\documentstyle[11pt,paspconf,psfig]{article}

\begin{document}

\title{Structure and Dynamics of Magnetized Interstellar
Clouds: Super-Alfv\'enic Turbulence?}

\author{Mordecai-Mark Mac Low}
\affil{Max-Planck-Institut f\"ur Astronomie, K\"onigstuhl 17, D-69117
Heidelberg, Germany}




\begin{abstract}
This review summarizes the argument for molecular clouds being
dominated by turbulence, most likely super-Alfv\'enic turbulence.
Five lines of observational evidence are given: molecular linewidths
and line shapes, nonequilibrium chemical abundances, fractal cloud
shapes, measurements of the dispersion of dust extinction through
clouds, and Zeeman measurements of the field strength.  Recent models
by Padoan \& Nordlund are summarized, that show that super-Alfv\'enic
turbulence appears consistent with these observations.  I then present
recent computations of my own, with numerical resolution as high as
$256^3$ zones, confirming the basic picture proposed by Padoan \&
Nordlund, but showing that the decay timescales they quote are
actually too long.  My computations show that decaying turbulence
loses its kinetic energy with a $\sim 1/t$ dependence on time
regardless of whether the turbulence is subsonic, supersonic,
isothermal, adiabatic, unmagnetized or magnetized.  Finally, the
implications for star formation and turbulence theory are discussed.
\end{abstract}




\section{Introduction}

Astronomers have understood qualitatively for at least half a century
that the interstellar gas may be in a turbulent state ({\em e.\ g.\ }
von Weizs\"acker 1951).  Coming to grips with this understanding in
quantitative models nonetheless remains an unfinished task, though
not for lack of attempts.  Rather, until very recently, most models
have had to be, perforce, fully analytic, without guidance from either
laboratory or numerical work.  Interstellar gas is magnetized,
strongly compressible, low viscosity, and has rms velocities far
higher than its thermal sound speed.  Achieving these conditions in
terrestrial laboratories would require extreme measures: even fusion
reactors are so small that kinetic instabilities dominate the
dynamics.  Three-dimensional numerical models with anything
approaching adequate resolution have only become possible in this
decade: the first transsonic, hydrodynamical model with 256$^3$ grid
points, still neglecting magnetic fields, was published by Porter,
Pouquet, \& Woodward in 1992, and no magnetohydrodynamic (MHD) model
with this resolution has yet appeared in the refereed literature,
although I will present some preliminary results from such models in
this paper, that I hope to have submitted before this paper is published.

The best observed regions of interstellar gas are molecular clouds
such as the Orion Molecular Cloud, as their high density allows the
formation of CO and other emitting species that can be followed in
great detail in mm-wave observations.  Molecular clouds appear
distinctly clumpy in both position and velocity space, leading to many
attempts to characterize them in terms of discrete clumps in a uniform
background medium, a viewpoint that can lead to misleading results, as I
will show below.  The temperatures inferred from line ratios suggest
that the observed line widths reflect motions far faster than the
local sound speed.  Whether these hypersonic velocities produce shocks
or not depends on the strength of the magnetic field, which has only
been determined in a very small number of regions in molecular clouds,
and so remains quite uncertain.  A preprint by Padoan \& Nordlund
(1997; hereafter PN) that I will discuss in this paper puts forward a
persuasive argument for the turbulence actually being super-Alfv\'enic as
well as supersonic.

The lifetimes of molecular clouds have been inferred to be several
$\times 10^7$ yr from the total fraction of gas mass in the Galaxy in
the form of molecular gas, and from the lifetimes of the young stars
associated with them (Blitz \& Shu 1980).  The argument has been made
that shock formation due to the observed hypersonic velocities would
dissipate the energy of the clouds so quickly that the entire clouds
would collapse and form stars in a time not much longer than their
free-fall times,
\begin{equation} t_{ff} = (1.4 \times 10^6 \mbox{ yr})(2 n(\mbox{H}_2) / 10^3
\mbox{ cm}^{-3})^{-1/2}, \end{equation} 
where $n(\mbox{H}_2)$ is the number density of molecular hydrogen
(Goldreich \& Kwan 1974, Field 1978).   

Three approaches have been taken to try to solve this problem.  One,
first taken by Arons \& Max (1975), is to argue that strong enough
magnetic fields will prevent shocks from occurring, and so lengthen
the dissipation time.  The second, taken by, for example, Scalo \&
Pumphrey (1982) is to argue, on the basis of clump models, that
hydrodynamic turbulence will actually dissipate more slowly than
expected.  Finally, at Schlo\ss\ Ringberg, I presented results from PN
that showed that MHD turbulence decays nearly as fast as
hydrodynamical turbulence.  Since then I have done models confirming
that result in principle, but showing that all sorts of turbulence
appear to decay even faster than they reported, with a time dependence
proportional to $\sim 1/t$.  If turbulence supports molecular clouds
against star formation, it must be constantly driven, by stellar
outflows (e.\ g.\ Silk \& Norman 1980), photoionization (McKee 1989,
Bertoldi \& McKee 1996), galactic shear (Fleck 1981), or some
combination of these or other sources.

\section{Review of Molecular Cloud Models}

Broadly speaking, four different descriptions of molecular cloud
structure have been proposed.  The first, and most straightforward, is
that the clouds consist of discrete clumps travelling on ballistic
orbits under the influence of gravity, colliding occasionally with
other clumps.  A second description, really a more sophisticated
version of the first, has the density distributed in a fractal
distribution, but still focusses on self-gravity as the dominant
force.  The third description invokes subsonic or sub-Alfv\'enic
turbulence, while finally, recent models have invoked supersonic,
super-Alfv\'enic turbulence.

The simplest models of clumpy, turbulent clouds describe the clouds as
made up of a large number of spherical gas fragments moving through a
lower-density surrounding medium (Scalo \& Pumphrey 1982),
possibly threaded by a magnetic field (Elmegreen 1985).  Scalo \&
Pumphrey showed that if all collisions were completely inelastic, with
any two fragments that came into contact sticking and dissipating
their relative energy, the energy would be dissipated on order of a
dynamical time $t_D = R/\langle v\rangle$, where R is a typical cloud size and
$\langle v\rangle$ is the root mean square velocity in the cloud.  If the turbulent
velocities are fast enough to support the cloud against gravitational
collapse, then the resulting estimated dynamical time is just of order
the free-fall time $t_{ff}$ (Field 1978).  It appears now that this is, in
fact, a reasonable estimate of how fast turbulent energy will be
dissipated. 

However, the picture of turbulence consisting of isolated spherical
clumps, when taken literally, has led to worse estimates of the
dissipation time scale.  For example, Scalo \& Pumphrey (1982) then
tried to extend their model by taking the geometry of the cloud
collisions into account, noting that off-center collisions of
spherical clouds would tend not to dissipate all of their energy.  The
outer parts of each cloud would simply slide by without being strongly
influenced by the impact.  This reduced the energy dissipation in
their model by an order of magnitude, bringing it into rough agreement
with cloud lifetimes, but not with more detailed models of turbulence
as I discuss below.  Elmegreen (1985) included magnetic fields in an
otherwise similar model, again reaching the conclusion that
dissipation could be much reduced, again in disagreement with more
realistic models of magnetized turbulence.  The fundamental flaw of
such models appears to be the neglect of the space-filling character
of even supersonic turbulence, which leads to rapid dissipation of the
turbulent energy.

This space-filling character can be described as a fractal structure.
Subsonic, incompressible turbulence is measured to have fractal
dimension $D = 2.3$ (Sreenivasan 1991, also see Sreenivasan \& Antonia
1997).  That is, measuring the volume of some tracer at different
resolutions, the measured volume changes as the resolution changes, as
if the tracer occupied a space of dimension intermediate between 2 and
3.  It remains unclear whether supersonic, magnetized turbulence has
the same fractal dimension, or indeed whether it has constant fractal
dimension at all (Chappell \& Scalo 1997).  Certainly cloud catalogs
have been used to derive a fractal dimension $D \sim 2.3$ (Elmegreen
\& Falgarone 1994).  An analytic model of an isothermal molecular
cloud that includes just the fractal density distribution and
self-gravity has been presented by Pfenniger \& Combes (1994), while
numerical models are presented by Klessen, Burkert, \& Bodenheimer (or
some permutation thereof) in these proceedings.

Strong magnetic fields are certainly observed in masers in the densest
regions of molecular clouds, and simple flux-conservation arguments
suggest that the interstellar field of 3--5~$\mu G$ should be
compressed to tens of $\mu G$ in large regions of molecular clouds.
Arons \& Max (1975) first suggested that MHD waves might drive the
observed supersonic motions.  Zweibel \& Josafatsson (1983) computed
the decay rates of such waves and found, in fact, that they decayed
within a free-fall time.  McKee \& Zweibel (1995) and Zweibel \& McKee
(1995) have shown that as long as such MHD waves are strong, they can
support clouds against gravitational collapse.  They make the
additional prediction that, in that case, the magnetic fields should
be in equipartition with the kinetic energy of the gas motions.
Gammie \& Ostriker (1996) have done 1D numerical models of MHD wave
support and dissipation.  They find relatively slow dissipation rates
for the waves, but our 3D models discussed below suggest that this is
due to their imposed symmetry.  We have reproduced their results in
1D, but find that in 1D, travelling shocks tend to combine, reducing
the dissipation rate, while in 3D, oblique collisions constantly
produce vorticity and new shocks.

Attempts to directly detect fields in molecular clouds using OH Zeeman
measurements have met with surprisingly limited success, however
(Troland et al.\ 1996), suggesting that magnetic fields may not be as
strong as expected.  Furthermore, the observed clumpy, possibly
fractal, density structure of the clouds is hard to produce with MHD
waves.  They will only be important if the magnetic energy density
exceeds the kinetic energy density.  However, in that case, the field
will adopt a simple geometry, tending to unfold any tangles or kinks
in the field lines.  Density enhancements will tend to form sheets
perpendicular to the field lines, but not isolated clumps (PN).  This
leads to the suggestion that super-Alfv\'enic motions may dominate the
structure of molecular clouds, producing structure fairly similar to
that seen in simulations of supersonic, unmagnetized turbulence.

\section{Observational Evidence}

There are at least five lines of observational evidence that support
the description of molecular clouds being at least turbulent, and
probably supersonically turbulent.  The most important is, of course,
the dynamical information gleaned from observations of tracer
molecules, especially isotopes of CO chosen to be optically thin in
the regions observed.  Second, examination of the chemistry required to
produce the observed abundances of molecules suggests that the
individual clumps are out of chemical equilibrium, and must therefore
be relatively young.  Third, the boundaries of clouds observed in the
infrared or in CO emission have fractal properties similar to those of
incompressible turbulent flows.  Fourth, recent measurements of the
dispersion of dust extinction through dark clouds can be naturally
reproduced by supersonic turbulent models.  Finally, Zeeman
measurements of magnetic fields yield values low enough that the
observed linewidths correspond to super-Alfv\'enic motions.

The first clue to the dynamics of molecular clouds is of course the
supersonic widths observed in CO lines.  I will not attempt to give a
full review of these observations, but merely mention two of the major
results that are directly relevant.  First, Larson (1981) described
correlations between linewidth, size, and density of clouds that have
guided much subsequent research.  Taken at face value, these relations
lead to the conclusion that the effective equation of state for the
gas in molecular clouds is logotropic (Lizano \& Shu 1989), a result
that does not agree with turbulent models of the molecular cloud gas
(V\'azquez-Semadeni, Cant\'o, \& Lizano 1997).  However, this may
reflect the limited validity of Larson's Laws, rather than the state
of the gas (see, for example, Scalo [1990] or V\'azquez-Semadeni,
Ballesteros-Paredes, \& Rodr\'{\i}guez [1997]).  Second, the
non-Gaussian shapes of the observed lines have been used to compare
models of turbulence with the observations (Falgarone \& Phillips
1990).  Turbulence is thought to produce intermittency that in turn
will produce such non-Gaussian lineshapes, specifically with more
power in the wings of the lines than would be expected in a Gaussian.

It has been known for some time that attempts to model the equilibrium
chemistry of molecular cloud cores yielded the puzzling result that
the early time ($10^5$--$10^6$~yr) nonequilibrium results appeared
more consistent with the observations than the final equilibrium
state.  A first attempt to understand this was made by Prasad, Heere,
\& Tarafdar (1991), who used a simple model of core collapse with
varying field strengths to follow the combined evolution of the
chemistry and the dynamics.  Xie, Allen, \& Langer (1995) used a
mixing-length description of diffusion to try to model chemistry in a
turbulent cloud, finding a better match to observations when
significant turbulence was included.  Bergin et al.\ (1997) use
observations of three cloud cores to come to conclusions similar to
Prasad et al.\ (1991): that the observed chemistry can only be
explained by cores with ages of order $10^5$ yr.  Such young cores are
a natural consequence of a turbulent flow in the presence of a
background radiation field that keeps low density material in a state
typical of the diffuse ISM.  Then the only material that will begin to
evolve to high-density, shielded, equilibrium states is material that
has been swept up in short-lived, turbulent clumps.

Attempts to describe the shapes of molecular clouds at different
scales led to the discovery that their boundaries in total column
density behave as fractals with a fractal dimension of $D \sim$
1.3--1.4 (Beech 1987, Dickman, Margulis, \& Horvath 1990).  Careful
examination of the edges of molecular clouds taking into account their
velocity structure using high-resolution CO observations by Falgarone,
Phillips, \& Walker (1991) confirmed this result.  Sreenivasan (1991)
presents an extensive review of results from terrestrial observations
of incompressible hydrodynamical turbulence that shows this dimension
to be typical of two-dimensional slices through turbulence.  While
these results are very suggestive, the relationship between
incompressible turbulence and the compressible, magnetized turbulence
presumably characteristic of molecular clouds remains uncertain, as
does the relationship between two-dimensional slices, and projections
of the full three-dimensional distribution into a position-velocity
cube. 

\begin{figure}
\psfig{file=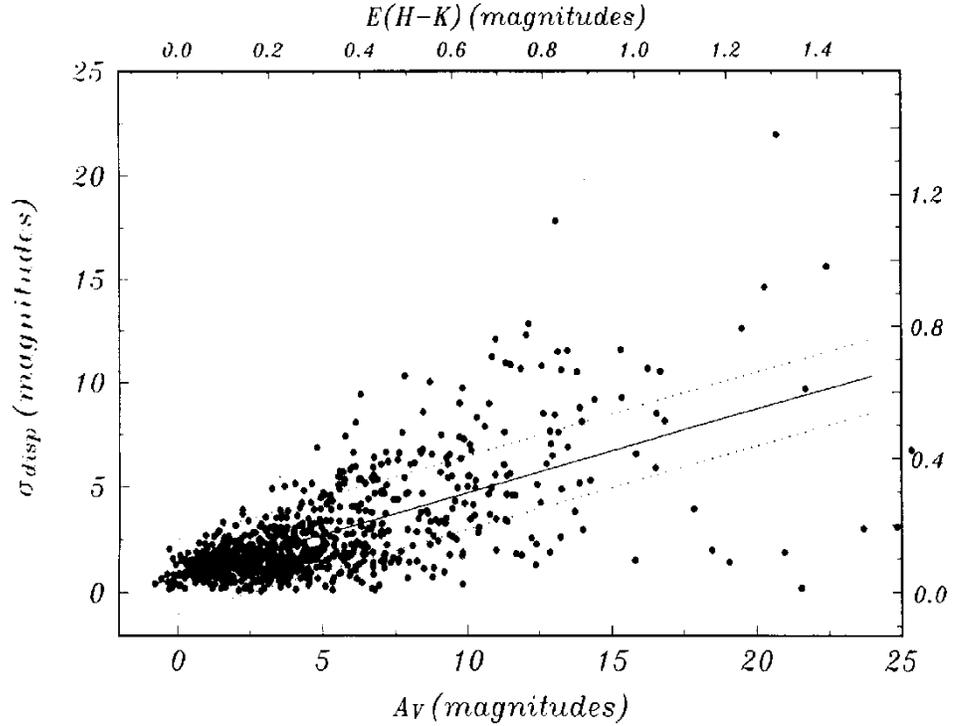,angle=270,width=\hsize,clip=}
\caption[junk]{Comparison of $\sigma_{disp}$, the dispersion in extinction
measure across each box, with $A_V$, the mean extinction in each box,
for the dark cloud IC~5146 (Lada et al.\ 1994).  Also plotted is the
linear, least-squares fit to the data.} \label{ladaobs}
\end{figure}

\begin{figure}
\psfig{file=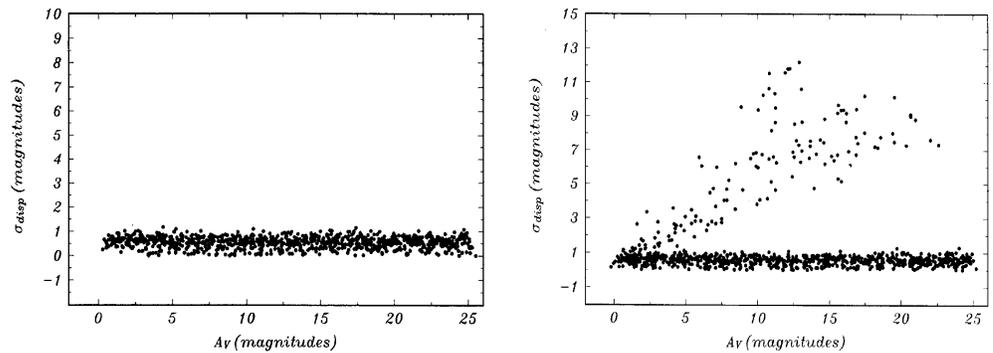,angle=0,width=\hsize,clip=}
\caption[junk]{Two simple models of dispersion in extinction versus mean
extinction from Lada et al.\ (1994).  In the left model, extinction
varies from box to box, but is uniform in each box, while in the right
model foreground stars are added to the stellar population used as
sample points.} \label{ladamod}
\end{figure}

Background stars can be observed in the near infrared through regions
with optical extinction as high as 30 magnitudes.  By observing the
infrared colors of these stars, their dust reddening can be measured,
and so the column density along the line of sight to the star can be
derived.  Lada et al.\ (1994) developed this method and applied it to
the dark cloud IC~5146.  They binned the region into boxes of size 1.5' in
order to measure the mean extinction, and found the very interesting
correlation shown in Figure~\ref{ladaobs}: the dispersion in the
extinction increases roughly linearly with the mean value.  They
showed that simple models of cloud structure such as uniform density
within each box or isolated clumps drastically failed to reproduce
this observation, as shown in Figure~\ref{ladamod}, but that they
could reproduce the observations with
a power-law variation of column density across each box.  Although
this column density distribution is quite artificial, a log-normal
probability distribution function of density, possibly characteristic
of isothermal, supersonic turbulence, has also been shown to reproduce
the observed correlation by Padoan, Jones, \& Nordlund (1997b), while
PN have used direct numerical simulation to reproduce the observation
with a weak-field MHD computation, as I will discuss below.

Measurements of the Zeeman effect in OH emission lines from molecular
clouds by Crutcher et al.\ (1993) and Troland et al.\ (1996) reveal
surprisingly low average field strengths $|\vec{B}|_{\parallel} \leq 10
\mu$G.  If the fields were in equipartition with the kinetic energy,
however, the field strengths should be of order 100 $\mu$G (Troland et
al.\ 1996).  Several different explanations are suggested by Troland
et al., including the possibility that the clouds really are in the
state of super-Alfv\'enic turbulence that a naive interpretation of the
results would suggest.  They point out however, that unfavorable
geometries, tangled fields, or OH abundance changes at high densities
could account for the limited observations to date.

\section{Models of Supersonic Turbulence}

Full hydrodynamical models of molecular clouds were first performed in
2D by Passot, Pouquet \& Woodward (1988), showing that even transsonic
flows developed the filamentary density enhancements characteristic of
molecular clouds.  They progressed to 3D in Falgarone et al.\ (1994),
where they simulated observations of the $512^3$, transsonic,
adiabatic, decaying turbulence simulations of Porter, Pouquet, \&
Woodward (1994).  Models of strongly supersonic, driven turbulence
with an isothermal equation of state are briefly referred to in
Padoan, Nordlund, \& Jones (1997d), although full description is there
deferred to a paper still in preparation by Nordlund \& Padoan, so it
remains unclear how the additional physics actually improves the fit
to the observations.

The most important new piece of physics, though, is the inclusion of
magnetic fields.  Stone (1995; also see Ostriker 1997) and Balsara,
Crutcher, \& Pouquet (1997) have published preliminary results from 3D
models, while Gammie \& Ostriker (1996) thoroughly modelled a
turbulent, strongly magnetized, molecular cloud in 1D.  PN have made a
major advance in the field by performing 3D MHD simulations of
turbulence in both strongly magnetized and weakly magnetized clouds.

PN used an Eulerian MHD code to model decaying, isothermal turbulence
in a box with periodic boundary conditions at a numerical resolution
of $128^3$ grid points.  Their code uses a staggered grid; is fifth
order in space and third order in time; uses hyper-diffusive fluxes;
and resolves shocks with an artificial viscosity and current sheets
with an artifical resistivity.  Although the high order of the code is
an advantage for smooth flows, eliminating diffusivity except at
scales very close to the grid scale, the advantage is lost in flows
with shocks or other discontinuities.  The discontinuities are
resolved over several zones by the artificial viscosity or resistivity
just as they would be in a lower-order code; meanwhile the higher
order method greatly increases the complexity and computational cost
of the code.  I present below preliminary models done with ZEUS-3D,
which is only second order in space and first order in time, but that
use $256^3$ zones, or eight times the number of zones of the PN
results.

Nevertheless, the interpretation of the two runs presented in PN
forces the serious consideration of their suggestion that the
turbulent motions in molecular clouds are actually strongly
super-Alfv\'enic.  The two runs are started with a solenoidal
perturbation of the velocity with root-mean-square (rms) amplitude of Mach 5
and maximum wavenumber of two---that is, the initial perturbations are
large and smooth, rather unlike the final turbulent state that has
perturbations at all scales.  The justification for this is that they
believe the driving forces to be from galactic shear, and so coming
from large scales initially (Nordlund, private comm., 1997).  Self-gravity
and ambipolar diffusion are neglected.  The field begins uniform and
vertical.

The strong field run has initial Alfv\'en number $A = \langle v\rangle
/v_A = 1$.  The initial velocity perturbations were set purely
perpendicular to the magnetic field to simulate an initial
distribution of Alfv\'en waves.  Although at the end Alfv\'en waves
remain the strongest component, motions parallel to the field have
increased to about half the velocity of the perpendicular motions.
Advection perpendicular to the field is strongly suppressed, resulting
in the formation of sheet-like high-density regions perpendicular to
the field.  The field remains close to uniform throughout the run, as
it is strong enough to resist tangling, as would be true for any field
in equipartition with the gas motions.

The weak field run has initial rms Alfv\'en number $A = 10$, with
uniform velocity perturbations.  Now the gas motions are strong enough
to overwhelm the field and it is swept along as the gas forms into the
clumps and filaments typical of supersonic hydrodynamic turbulence.
The field does dominate initially in the densest swept-up regions, but
even in them, advection along field lines can increase the
mass-to-flux ratioes significantly over time.  Field strengths and
directions vary greatly across the region, and the typical morphology is
more filamentary or clumpy than sheet-like.

\begin{figure}
\psfig{file=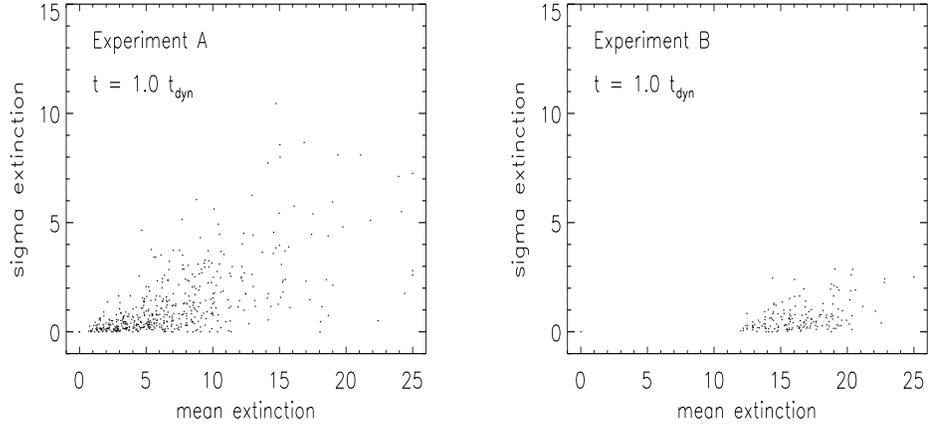,angle=0,width= \hsize,clip=} 
\caption[junk]{Comparison between dispersion of extinction values and mean
extinction values in regions across the numerical models presented by
PN. Experiment A is the weak field model,
which reproduces the observations shown in Figure~\ref{ladaobs}
significantly better than Experiment B, the strong field model.  The
values of the extinction were measured at random locations siumulating
the random position of a star behind the cloud.} \label{pndisp}
\end{figure}

\begin{figure}
\psfig{file=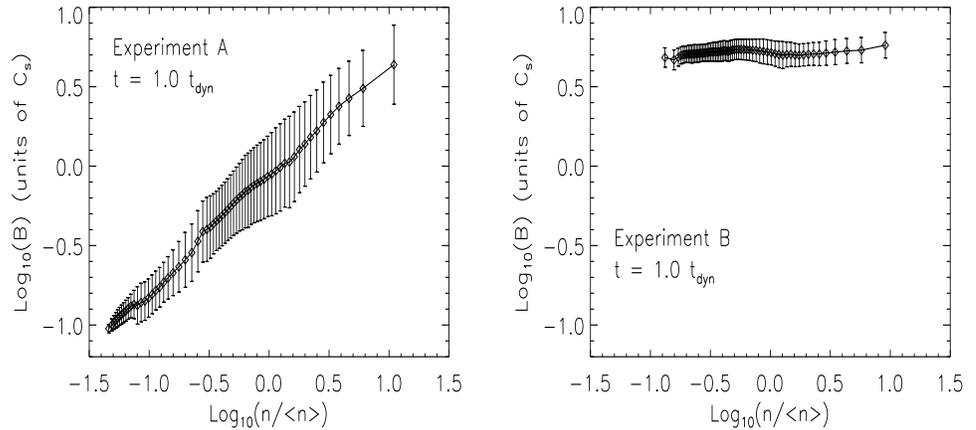,angle=0,width=\hsize,clip=}
\caption[junk]{Magnetic field strength versus density after one
dynamical time in the weak ({\em left}) and strong ({\em right}) field
numerical models of PN, showing that the weak field model reproduces
the observed variation of field with density, while the strong field
model has the same field strength regardless of local density
everywhere.  The magnetic field is expressed in units such that
$\vec{B} = 1$ implies $v_A = 1$ and $c_s = 1$ for the mean density
$\langle n\rangle$.  The bars show the $1\sigma$ dispersion of values
in each density bin.} \label{pnbn}
\end{figure}

These clear differences between the weak and strong field runs can be
compared to the observations.  The weak field run reproduces the
observations better than the strong field run in at least three ways.
First, the clumpy morphology matches the morphology observed in, for
example, CO maps.  Second, the clumpiness produces column density
dispersions that match the Lada et al.\ (1994) results better than the
relatively uniform sheets of the strong-field run, as shown in
Figure~\ref{pndisp}.  
Third, the variation in magnetic field strength with density in the
weak field run is better able to reproduce the observed correspondence
of magnetic field with density than the strong field run, in which the
field remains uniform regardless of the density, as shown in
Figure~\ref{pnbn}.  This comes from the model very naturally, because
in the strong field case, the field is strong enough to resist
compression, and so maintains a roughly uniform strength everywhere,
while in the weak field case, the field is carried with the flow,
increasing in strength in the same regions the density increases.

Since the meeting, Padoan and his collaborators have submitted papers
in which they begin to directly compare their observations with
observations in molecular lines.  Padoan et al.\ (1997a) describes a
non-LTE radiative transfer computation that they apply to the results
of the PN computations to generate simulated observed spectra.  They
note that the resulting spectra and maps resemble the observations in:
morphology; intermittency in the wings of the spectra; smooth central
peaks in the integrated lines; multiple components along individual
lines of sight; statistical moment distributions; and the
linewidth-intensity relation.  Padoan et al.\ (1997c) make
comparisons between these simulated observations and actual observations
of the Perseus Molecular Cloud, drawing similar conclusions.

\begin{figure}
\psfig{file=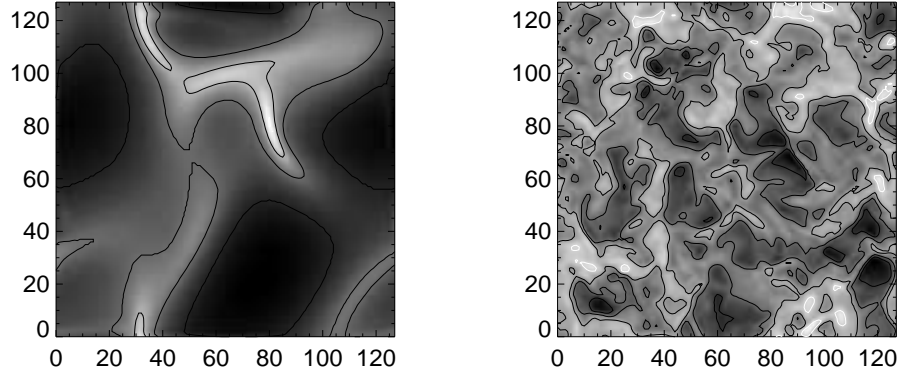,angle=0,width=\hsize,clip=} 
\caption[junk]{Comparison of log density for slices through Mach 5 decaying
turbulence started with different initial conditions, at a time of
$10^{-2} L/c_s$, where $L$ is the box size and $c_s$ the sound speed.
Initial conditions included only wavenumbers up to two on the left,
while the initial conditions on the right included wavenumbers up to
eight, demonstrating the long transition period required for a
low-wavenumber initial condition to reach a fully turbulent state.
These computations were performed on a $128^3$ grid using ZEUS-3D.
The same greyscale is used for both images, covering two orders of
magnitude in density.} \label{kmax}
\end{figure}

\begin{figure}
\psfig{file=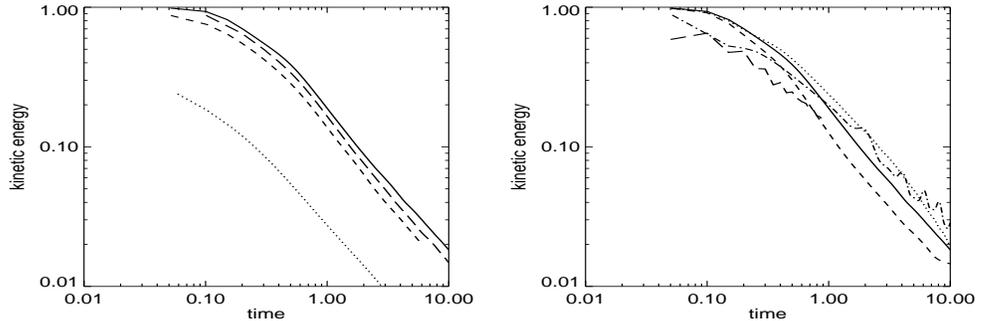,angle=90,width=\hsize,height=0.5\vsize,clip=}
\caption[junk]{({\em left}) Resolution study for decaying, supersonic,
isothermal, hydrodynamic turbulence with initial rms Mach number 5,
and numerical resolutions of $32^3$ ({\em dotted}), $64^3$ ({\em short
dashed}), $128^3$ ({\em long dashed}), and $256^3$ ({\em solid})
zones.  Time is given in units with $L/c_s = 20$, where $L=2.0$ is the
size of the box and $c_s = 0.1$ is the sound speed.  The initial
conditions included perturbations up to a maximum wavenumber of eight.

({\em right}) Comparison of decay of kinetic energy for models with
different physics at a resolution of $256^3$ zones and the same parameters
as above: ({\em solid}) isothermal hydrodynamics, ({\em dotted})
adiabatic hydrodynamics, with adiabatic index $\gamma = 1.4$, ({\em
short dashed}) weakly magnetized, with initial rms Alfv\'en number $A
= 5$, ({\em dot-dashed}) moderately magnetized, with $A = 1$, and
({\em long dashed}) strongly magnetized, with $A = 0.05$.  All lose
kinetic energy proportional to $t^{-1}$, similar to the well-resolved
hydrodynamic model. 
} \label{resfig}
\end{figure}

I have now reproduced and extended the PN computations myself, using
ZEUS (Stone \& Norman 1992a, b), and find the same morphology, as
described above.  However, I find that their decay timescales are
strongly dependent on their very smooth initial conditions.  Fully
developed supersonic turbulence actually appears to decay
significantly faster than they claim, whether or not it is dominated
by magnetic fields.  The problem is that most of the time covered by
the computations reported in their paper is taken up by the transition
from their smooth initial condition to a turbulent state, so that the
actual behavior of the turbulence is only seen at the very end of
their runs.  In Figure~\ref{kmax} I compare the density distribution
shortly after start for a run with maximum wavenumber two, as used by
PN, and a run with maximum wavenumber eight.  Eventually both runs
reach equivalent states, but the run with low wavenumber takes several
dynamical times to do so; this initial transient should not be taken
as representative of turbulent behavior.  We do agree with their
ultimate conclusion that the inclusion of magnetic fields does not
greatly change the decay timescale of the turbulence, as shown in the
right panel of Figure~\ref{resfig}, but disagree on what that
timescale is.

Our computations have, rather surprisingly, revealed this timescale to
be quite universal: undriven turbulence decays as $t^{-\alpha}$, with
$\alpha$ in the range $1 < \alpha < 1.4$, whether the turbulence is subsonic,
supersonic, hydrodynamic, magnetically dominated, isothermal or
adiabatic, as demonstarted in Figure~\ref{resfig}.  This result is
very well converged as demonstrated in the isothermal, hydrodynamical
case in the left panel of this Figure.
This result is also consistent with experimental measurements
(Comte-Bellot \& Corrsin 1966, Warhaft \& Lumley 1978) and theoretical
models (Lesieur 1997) of incompressible turbulence.  It is
nevertheless quite unexpected, since those theoretical models relied
heavily on the specific properties of incompressible hydrodynamic
turbulence, rather than universal properties extending across all the
types of turbulence we have now simulated.

\section{Implications}

\subsection{Star Formation}


If molecular clouds are in a state of driven, super-Alfv\'enic
turbulence, I would outline a scenario for star formation that
addresses a number of questions that remained unanswered in the
standard scenario of star-formation primarily mediated by ambipolar
diffusion in quasi-static cores ({\em e.\ g.\ } Mouschovias 1991).  One of
the reasons that the standard scenario was proposed was to solve the
problem of explaining why molecular clouds did not collapse into stars
within a single free-fall time.

Under the new scenario, the clouds are supported against collapse by
turbulence driven either from galactic shear or local stellar energy
sources.  Periodically the violent density fluctuations produced by
super-Alfv\'enic shocks will form clumps greater than the {\em local}
Jeans mass, which will then begin to gravitationally collapse until they
are supported by their magnetic fields.  If the gravitational binding of
these clumps is sufficiently strong, they will no longer be influenced
by the surrounding turbulent flow, and can thereafter evolve as
described by the standard scenario, producing low mass stars.

Rarely, however, the turbulent converging flows will also produce
density concentrations much larger than a local Jeans mass that are
immediately magnetically supercritical.  Accretion down field lines
driven by the external velocity field can raise the central
mass-to-flux ratioes of these clumps high enough to collapse without
going through a quasi-static phase.  This would produce at least
intermediate mass stars and perhaps even high mass stars, although
there the question of how the countervailing radiation pressure is
overcome remains open.

Because they form in a turbulent medium, a wide range of rotational
velocities would be expected for both small and large cores.  This
would lead to a wide range of fragmentation behavior ({\em e.\ g.\ }
Burkert \& Bodenheimer 1996), and a stellar initial mass function that
can not be simply derived from from the observed core or clump initial
mass function.  I note that the attempt by Padoan et al.\ (1997d) to
derive an initial mass function (IMF) from the probability density
function of density in supersonic turbulence has been strongly
criticized for this and other reasons by Scalo et al.\ (1997).  My
personal suspicion is that Adams \& Fatuzzo (1996) must come closer to
the truth in trying to describe the initial mass function as the
result of many random variables operating together.  The turbulent
production of the parent cores might provide much or all of the
necessary randomization.  However, they, too, end up with a log-normal
IMF that does not necessarily agree well with the observations (Scalo
et al. 1997).  Sreenivasan (1991, p. 592) points out that the failing
assumption here is that all the processes will be distributed so that
the central-limit theorem holds, and that turbulent intermittance
produces rare but large events that do not follow that theorem.  He
suggests that a multifractal formalism might provide a way forward, as
has now begun to be explored by Chappell \& Scalo (1997).

\subsection{Turbulence}

Our new computations, presented in Figures~\ref{resfig}, showing that
undriven turbulence decays as approximately $1/t$,  emphasize
that such turbulence will lose its kinetic energy quickly,
regardless of the details of the state of the gas.  As the gas in
molecular clouds is observed to have significant kinetic energy, that
kinetic energy must be supplied from somewhere on a more or less
continuous basis.   If turbulence supports molecular clouds
against star formation, it must be constantly driven, by stellar
outflows (e.\ g.\ Silk \& Norman 1980), photoionization (McKee 1989,
Bertoldi \& McKee 1996), galactic shear (Fleck 1981), or some
combination of these or other sources.

Our computations also suggest that significant progress must be
possible in the theory of turbulence, as we have uncovered a
general behavior that does not depend on the details of the cascade of
energy down to the dissipative scale.  Our results must, of course, be
compared to experiment, and verified with more sophisticated numerical
models---our models do not, for example, include an explicit model for
diffusivity, but merely rely on numerical diffusivity to diffuse
energy at the smallest scales.  The excellent convergence shown in the
first panel of Figure~\ref{resfig} suggests, however, that as is
usually assumed in classical incompressible turbulence theory, the
details of the dissipative process matter rather less than its
presence only at the smallest scales.


\acknowledgments Original work presented in this paper was done in
collaboration with R. Klessen, A. Burkert, and M. D. Smith.
Computations were performed at the Rechenzentrum Garching of the
Max-Planck-Gesellschaft.  I thank E. Vazquez-Semadeni for interesting
discussions of turbulence, and P. Padoan for generously allowing the use
of figures from his work prior to publication.  In preparation of this
review I have made extensive use of NASA's Astrophysics Data System
Abstract Service.

\end{document}